\documentstyle[12pt,psfig]{article}
 
\font\blackboard=msbm10 at 12pt
\font\blackboards=msbm7
\font\blackboardss=msbm5
\newfam\black
\textfont\black=\blackboard
\scriptfont\black=\blackboards
\scriptscriptfont\black=\blackboardss

\newcommand{\tr}{{\rm Tr}}

\newcommand{\lag}{{\cal L}}

\newcommand{\gone}[1]{}
\begin{document}
\pagestyle{plain}
\setcounter{page}{1}
 
\baselineskip16pt
 
\begin{titlepage}
 
\begin{flushright}
NSF-ITP-97-133\\
PUPT-1737\\
hep-th/9710240\\
\end{flushright}
\vspace{20 mm}

\begin{center}

{\Large {\bf The Spectrum of a Large N Gauge Theory
Near Transition from Confinement to Screening
}}

\bigskip

\centerline{David J.\ Gross$^a$, Akikazu Hashimoto$^a$, 
and Igor R.\ Klebanov$^b$}

\bigskip

\begin{center}
{\it $^a$Institute for Theoretical Physics \\
University of California\\
Santa Barbara, CA 93106-4030\\

\bigskip

$^b$Department of Physics\\
Joseph Henry Laboratories\\
Princeton University\\
Princeton, New Jersey 08544-0708
}
\end{center}

\end{center}
\begin{abstract}
\noindent We study the spectrum of 1+1 dimensional large $N$ QCD
coupled to an adjoint Majorana fermion of mass $m$. As $m\rightarrow
0$ this model makes a transition from confinement to screening. We
argue that in this limit the spectrum becomes continuous for mass
greater than twice the mass of the lightest bound state. This critical
mass is nothing but the threshold for a decay into two lightest
states. We present numerical results based on DLCQ that appear to
support our claim.
\end{abstract}

\begin{flushleft}
October 1997
\end{flushleft}
\end{titlepage}
\newpage

\section{Introduction}

Deriving the low-energy properties of QCD continues to be a largely
unsolved problem. For this reason one often resorts to simpler
low-dimensional models in order to gain intuition about the $3+1$
dimensional case. A classic such model is the `t Hooft model
\cite{GT}, which is the $1+1$ dimensional $SU(N)$ gauge theory coupled
to Dirac fermions in the fundamental representation.  Two key elements
in the solution of the model are the large $N$ limit and the
light-cone quantization.  The large $N$ limit simplifies the dynamics
by removing the interactions between confined states. In the `t Hooft
model one thus finds a single Regge trajectory of non-interacting
mesons.  Quantization on the lightcone, originally introduced in
\cite{Dirac49}, is a useful tool as well, because all unphysical
degrees of freedom become manifestly non-dynamical and can be
eliminated using the constraints.

In light cone quantization, one treats one of the null co-ordinates,
usually chosen to be $x^+$, as the time. The other null coordinate is
then treated as spatial, and we could imagine compactifying it into a
circle. Then the spectrum of the longitudinal momentum $p_-$ becomes
discrete, hence the name Discrete Light Cone Quantization (DLCQ)
\cite{masyama75,PauliBrodsky85}. This approach to theories on the light-cone is
often useful as a conceptual tool (as, for example, in the Matrix
formulation of M-theory \cite{susskind97}). It is also a practical
device for solving theories numerically.  By now there is an enormous
literature on QCD on the light cone. Readers are referred, for
example, to \cite{BPP97} for a recent review and list of references.

While the `t Hooft model provides the simplest demonstration of
confinement in a non-abelian gauge theory, it does not capture the
complexity of $3+1$ dimensional gauge dynamics.  This is because it
contains no dynamical degrees of freedom in the adjoint representation
of $SU(N)$.  In order to model the physics of transverse gluons, one
may consider $1+1$ dimensional QCD coupled to matter in the adjoint
representation \cite{IgorAdjFermi}. Such degrees of freedom can be
thought of as arising from dimensional reduction of QCD in higher
dimensions. A particular model which has received some attention
recently \cite{IgorAdjFermi,kutasovSUSY,BDK} is that of a single
Majorana fermion in the adjoint representation coupled to
two-dimensional QCD,
\begin{equation}
S = \int d^2 x \tr \left( i \bar{\Psi}^T {D\!\!\!\!/\,}  
\Psi - m \bar{\Psi}^T \Psi - 
{1 \over 4g^2} F_{\alpha \beta} F^{\alpha \beta}\right).
\label{AdjFermiModel}
\end{equation}
In many ways, this is the simplest model exhibiting some interesting
physical features. This theory is manifestly finite (unlike the model
coupled to an adjoint scalar which requires a mass renormalization)
and its numerical investigation can be easily set up using the
Discrete Lightcone Quantization
\cite{IgorAdjFermi,BDK,KresimirIgor,AntonPinsky96}.  The model
contains one adjustable dimensionless parameter $x = {\pi m^2 \over
g^2 N}$.  Several interesting features of this model have been noted
in the literature.  For example, at the special value of the
parameter, $x=1$, the model becomes supersymmetric \cite{kutasovSUSY}.
Unlike the `t Hooft model, the theory (\ref{AdjFermiModel}) has an
exponentially increasing density of bound states, $\rho(M) \sim
e^{M/T_H}$ \cite{kutasovSUSY,BDK}.  Thus, at temperature $T_H$ a
deconfinement transition occurs.  Surprisingly, the temperature $T_H$
exhibits a non-trivial dependence on mass; for example,
$T_H\rightarrow 0$ as $m\rightarrow 0$.  This is because in the
massless limit the theory undergoes a phase-transition from the
confining phase to the screening phase \cite{GKMS96,AFS97}.  The
string tension scales linearly with the fermion mass and vanishes at
the point of the phase transition.

In order to improve our insight into the transition to screening, we
need to understand, at least qualitatively, what happens to the
spectrum of string-like bound states as $m\rightarrow 0$.  On the one
hand, for any $m>0$ the theory is confining, hence the spectrum is
strictly speaking discrete all the way to infinite mass. On the other
hand, in the limit $m\rightarrow 0$, one should expect that the
spectrum becomes continuous, at least above a certain mass. A
heuristic argument for this goes as follows \cite{GKMS96}.  In a
screening theory there is finite range attraction between color
non-singlets which may be strong enough to create a few bound states.
However, since the attractive potential flattens at infinity, we
expect the spectrum to be continuous above a certain mass.  In this
paper we present numerical results that, indeed, appear to be
consistent with this simple picture.

Our results are also consistent with some of the findings in
\cite{KutasovSchwimmer,BK97}. There it was argued that for
$m\rightarrow 0$ one can identify certain ``basic'' bound states
(single particles).  The spectrum of these particles is discrete.  For
small $m$ most of the string states may be thought of as loosely bound
multi-particle states. These multi-particle states form a continuum as
$m\rightarrow 0$.

The principal new result of this paper is that the continuum begins at
twice the mass of the lightest particle.  We believe that the spectrum
is discrete below this critical mass.  While this structure of the
$m\rightarrow 0$ limit of the spectrum could have been anticipated
from the arguments reviewed above, we support it by careful analysis
of the DLCQ numerical diagonalizations.

\section{Decoupling between massless and massive sectors}

Let us begin by briefly reviewing the arguments of
\cite{KutasovSchwimmer}. Consider a conformal field theory invariant
under global symmetry group $G$. Such a model arises as a
representation of affine Lie algebra $\hat{G}$. Consider, for example,
a Lagrangian for a theory with right handed quarks $\psi^{(r)}$ and
left handed quarks $\chi^{(r')}$ in representations $r$ and $r'$ of
$G$ respectively:
$$\lag_{CFT} = \sum_r \psi^{\dag (r)} 
\partial_+ \psi^{(r)} + \sum_{r'} \chi^{\dag (r')} \partial_- \chi^{r'}
\ .$$
There is a natural way to couple such a theory to a non-abelian gauge
field based on gauge group $G$:
$$\lag = \lag_{CFT} + A_{+a} J^{+a} + 
A_{-a} J^{-a} + {1 \over 2g^2} (\partial_- A_+ - \partial_+ A_-)^2 $$
with
\begin{eqnarray*}
J^{+a} & = & \psi^{\dag(r)} \lambda^{a (r)} \psi^{(r)}\ ,\\
J^{-a} & = & \chi^{\dag(r)} \lambda^{a (r)} \chi^{(r)}\ .
\end{eqnarray*}
This theory is believed to be consistent if the levels $k$ and
$\bar{k}$ of the left and right moving KM currents coincide.

In light cone quantization, one treats the light like coordinates
$x^-$ and $x^+$ as space and time, respectively.  The natural gauge in
this coordinate system is the light cone gauge $A_- = 0$. In this
gauge, $A_+$ and $\chi$ are non-dynamical. Taking into account the
constraints imposed by these non-dynamical fields, the light cone
Hamiltonian becomes
\begin{equation}
P^-  = -  \int dx^-  \frac{1}{2} g^2 J^+ {1 \over \partial_-^2} J^+.
\label{LightconeHamiltonian}
\end{equation}
The dependence on $\chi$'s has disappeared from the Hamiltonian, other
than the basic requirement that their chiral anomaly matches that of
the $\psi$'s.  This implies that $\chi$ could have been replaced with
any other representation of $\hat{G}$ as long as they have the same
chiral anomaly.

Strictly speaking, it is incorrect to conclude that the physics of
these models depends only on the KM level of the matter content,
because by going to light cone coordinates, the dynamics of massless
degrees of freedom propagating along the $x^-$ axis is lost.  This
implies, however, that all data specifying the details of matter
representation other than its KM level is encoded in the massless
sector propagating along $x^-$. The massless left-moving sector is
therefore decoupled from the massive sector.

However, instead of treating $x^-$ as space and $x^+$ as time, one
could have considered taking $x^+$ as space and $x^-$ as time. Then,
the natural gauge choice would have been $A_+=0$, and using the same
argument as the one given above, one concludes that the massless
right-moving sector is decoupled from the massive sector.

$SU(N)$ gauge theory coupled to a massless adjoint fermion introduced
in (\ref{AdjFermiModel}) is an example of such a gauged WZW model. The
current $J^{ab}=2 \psi^{ac} \psi^{cb}$ generates a KM algebra of level
$N$. For generic mass, the single particle color singlet states in
this model are of the form
\begin{equation}
\tr ( \psi \psi \ldots \psi) | 0 \rangle
\label{singlets}
\end{equation}
and these states becomes non-interacting in the $N \rightarrow \infty$
limit.

For $m=0$, however, the form of the light-cone Hamiltonian
(\ref{LightconeHamiltonian}) suggests that the Hilbert space of single
particle states can be block diagonalized into current blocks labeled
by the KM primaries.  The simplest states in the current blocks are of
the form
$$ \tr(J J J \ldots J ) | 0 \rangle$$
or
$$ \tr (J J J \ldots J \psi)| 0 \rangle\ .$$
General highest weight states are of the form
\begin{equation}
 \left(\prod_{i=1}^n \psi^{a_i b_i} \right) | 0 \rangle
\label{highestWeight}
\end{equation}
with symmetrization of indices $a_i$ and $b_i$ encoded in terms of
Young tableaux with $n$ boxes. A generic state in the current block
with $n$ fermions in the primary will be of the form
$$ \tr (J^{l_1} \psi J^{l_2} \psi \ldots J^{l_n} \psi)  |0 \rangle$$

$SU(N)$ gauge theory coupled to $N$ flavors of massless fundamental
fermions is also an example of a gauged level $N$ WZW model.  The
decoupling theorem implies that the physics in the massive sector
should agree with that of the adjoint fermion model described earlier.
The highest weight states of the KM algebra are of the form
$$ \left( \prod_{i=1}^n = \psi^{\dag a_i \alpha_i} 
\psi^{\dag b_i \beta_i} \right) |0 \rangle$$
where $a_i$ and $b_i$ are symmeterized just as in the adjoint fermion
case and are characterized by Young tableaux with $n$ boxes.  A
generic state in such a current block sector will be of the form
$$
\left[ \psi^{\alpha_1 a_1} (J^{l_1})^{a_1 b_1} \psi^{\dag b_1 \beta_1} \right]
\left[ \psi^{\alpha_2 a_2} (J^{l_2})^{a_2 b_2} \psi^{\dag b_2 \beta_2} \right]
\ldots
\left[ \psi^{\alpha_n a_n} (J^{l_n})^{a_n b_n} \psi^{\dag b_n \beta_n} \right] | 0 \rangle
$$
For $n \ge 1$, these states appear to correspond to $n$ mesons built
out of fundamental quarks.

In general, the states in the massive sector are labeled by the
current block and the currents. Based on this classification, a state
$$ | \Phi \rangle =\tr ( J^m \psi ) | 0 \rangle$$
of the adjoint fermion model is associated with a state
$$ | \Sigma \rangle =\left[ \psi^{\alpha a} 
(J^{m})^{a b} \psi^{\dag b \beta} \right] | 0 \rangle
$$
of the $N$-flavored fundamental model. There are two problems with
this identification. Firstly, $|\Phi \rangle$ is a fermion whereas
$|\Sigma\rangle$ is a boson. Secondly, $|\Phi\rangle$ is a unique
state whereas $|\Sigma\rangle$ has $N^2$-fold degeneracy due to choice
of flavors.  The resolution to this apparent discrepancy lies in the
massless sector we have ignored up till now. The state
$|\Sigma\rangle$ is actually $|\Phi \rangle \otimes | \Xi^{\alpha
\beta} \rangle$ where $|\Xi^{\alpha \beta}\rangle$ is a fermionic
state from the massless sector of the theory.  These states carry
flavor index, as is expected of the massless sector which contains all
information about the matter representation beyond its KM
level. Dynamically, these states simply act as spectators and are
otherwise decoupled.

Taking the massless sector into account, the structure of the full
Hilbert space is expected to be of the form
\begin{equation} \label{top}
{\cal H} = \sum_{s,s'} \oplus ( 
{\cal H}_s^c \otimes
{\cal H}_{s,s'}^{GI} \otimes
{\cal H}_{s'}^c 
)
\ .
\end{equation}
where $s$ and $s'$ labels the representation of the KM algebra on the
left and right movers, respectively.  $H_s^c$ encodes the massless
spectrum of the theory which is model dependent.

For the adjoint fermion there are no massless states (this is a
consequence of the fact that the central charge vanishes for the
infrared conformal field theory). Nevertheless, $H_s^c$ is
non-trivial: it is built of certain discrete topological states which,
roughly speaking, label the distinct vacua for each of the current
block sectors.  Thus, the spectrum of the adjoint fermion model in the
limit $m \rightarrow 0$ should exhibit
\begin{enumerate}
\item A set of ``basic'' bound states (particles) with masses $m_1$,
$m_2$, $\ldots$. They give rise to $n$-body thresholds at $m^2 =
(\sum_{j=1}^n m_{i_j})^2$. The lowest such threshold is the most
obvious because this is where the spectrum becomes continuous. It is
expected to occur at $m^2= 4 m_1^2$.
\item There should be degeneracy arising from the tensor product
structure (\ref{top}) of the physical states, which includes the
topological sector of the theory. One manifestation of it will be the
match between the continuous parts of the spectra of bosons and
fermions.
\end{enumerate}

In the following section, we will present numerical evidence based on
DLCQ for both of these features.

\section{Numerical Results from Discrete Light Cone Quantization}

\subsection{Review of adjoint fermion model in DLCQ}

The application of DLCQ techniques to the gauged adjoint fermion model
has been developed in
\cite{IgorAdjFermi,BDK,KresimirIgor,AntonPinsky96}.  We will briefly
review the construction below.

We start with the action (\ref{AdjFermiModel}), where $\Psi$ is an
adjoint Majorana fermion whose spinor components are given by $({ \psi
\atop \chi} )$. In lightcone coordinates and in the lightcone gauge,
(\ref{AdjFermiModel}) becomes
$$S = \int d x^+ \int d x^- \ (i \psi \partial_+ \psi + i \chi
\partial_- \chi - i \sqrt{2} m \chi \phi + { 1 \over 2 g^2}(\partial_-
A_+)^2 + A_+J^+)
\ .$$
Here, $J^+ = 2 \psi_{ik} \psi_{kj}$.  We see that $\chi$ and $A_+$ are
non-dynamical and simply lead to constraints.

In DLCQ, one compactifies the $x^-$ direction into a circle of period
$L$ and assign periodic boundary condition to the gauge fields. The
original two dimensional model is recovered in the decompactification
limit of this theory.  Having assigned periodic boundary condition for
the gauge fields, the equation of motion allows two possible boundary
conditions for the fermions: periodic or anti-periodic.  If periodic
boundary condition are used, the mode expansion of $\psi$ includes a
zero momentum component. It is customary to ignore this mode when
computing the DLCQ spectrum.  This is justified for generic $m$
because the constraint due to the zero-momentum component of $\chi$
will set the zero momentum component of $\psi$ to zero. At $m=0$,
however, this constraint disappears, and one is no longer justified in
throwing away the zero momentum component of $\psi$.  Unfortunately,
the simplest DLCQ cannot be applied in the presence of such a
zero-mode, as will be made clear shortly.  One could simply discard
the zero-mode (at least for $m>0$ this should not affect the spectrum
in the $K\rightarrow \infty$ limit).  The price we pay is that the
rate of convergence to the decompactification limit becomes much
slower.  This problem does not arise if one chooses to quantize the
fermions using anti-periodic boundary conditions, since the zero
momentum modes are absent from the beginning.  We have found
empirically that a DLCQ computation using anti-periodic boundary
conditions for the fermions indeed converges much faster compared to
the periodic boundary conditions.  Therefore, we will adopt the
anti-periodic boundary conditions as the method of choice, as did
\cite{BDK}.  It should be stressed, however, that the
decompactification in DLCQ is impossible to achieve in practice. Thus,
a careful extrapolation is needed to extract the features of the
spectrum alluded to at the end of the previous section.

In lightcone quantization, fermions are made to satisfy the canonical
anti-commutation relations imposed at equal lightcone time $x^+$:
$$\{ \psi_{ij}(x^-), \psi_{kl}(y^-) \} = 
{1 \over 2} \delta(x^- - y^-) (\delta_{il} \delta_{jk} - 
\frac{1}{N} \delta_{ij} \delta_{kl})\ .$$

In terms of the modes
$$
\psi_{ij}(x) = {1 \over \sqrt{2L}} \sum_{n \in {\rm odd}} 
B_{ij}(n) e^{-{\pi i n x \over L}}\ ,
$$
the anticommutation relations become
$$\{B_{ij}(m), B_{kl}(n)\} = \delta(m+n) (\delta_{il} \delta_{jk} - 
\frac{1}{N} \delta_{ij} \delta_{kl})\ ,$$
where $B_{ij}(-n)$ for $n > 0$ refers to $B_{ji}^{\dagger}(n)$ in the
notation of \cite{KresimirIgor}.

Taking the appropriate constraints into account, the lightcone
momentum and energy become\\
\parbox{\hsize}{\begin{eqnarray*}
P^+ & = & 
\sum_{n \ge 1} 
\left( {\pi n \over L} \right) B_{ij}(-n) B_{ij}(n)\ , \\
P^{-} &=&  
{m^2 \over 2}  
\sum_{n \ge 1} \left( {L \over \pi n} \right) B_{ij}(-n) B_{ij}(n) 
+ \sum_{n \ge 1} {g^2 L \over (\pi n)^2}  
J_{ij}(-n) J_{ij}(n)\ .\end{eqnarray*}}
Restricting to the sector where  $P^+ = \pi K / L$, we find
$$M^2 = 2 P^+ P^- = {g^2 N \over \pi}  K  \left( {m^2 \pi \over Ng^2}
\sum_{n \ge 1} {1 \over n} B_{ij}(-n) B_{ij}(n) + {1 \over N} \sum_{n
\ge 1} {1 \over n^2} J_{ij}(-n) J_{ij}(n) \right)\ .$$
We are interested in the $m\rightarrow 0$ limit of the spectrum.  For
any $m>0$ the use of DLCQ is completely justified, and we expect the
spectrum to converge as $K\rightarrow \infty$. Thus, we will imagine
taking $m$ very small in the formula above. If $m$ is small enough
(say $10^{-10}$), then our numerical calculations will not feel it at
all. Hence, we will simply set $m=0$ in the numerics, and think of the
extrapolation $K\rightarrow \infty$ as a representation of the
$m\rightarrow 0$ limit of the spectrum.

The Hilbert space on which the $M^2$ operator acts can be constructed
by acting on the vacuum with the mode operators $B(-n)$
$$ \tr (B(-n_1) B(-n_2) \ldots B(-n_l)) | 0 \rangle\ ,$$
subject to the condition that $\sum n_i = K$. These states are
generated by a set of ordered partitions of $K$ into odd integers, up
to graded cyclic permutations. There are only finitely many such
states. The $M^2$ matrix can be evaluated explicitly by commuting the
oscillators.  This is what makes DLCQ a powerful tool: the Hamiltonian
is a finite dimensional matrix which can be diagonalized numerically
(this feature breaks down in the presence of zero-momentum modes).

The decompactification limit of this theory is obtained by sending $L$
to infinity, keeping $P^+$ constant.  It is then necessary to scale
$K$ with $L$. This is exactly the sense in which the
decompactification limit is a challenging limit in DLCQ. In general,
the number of partitions of a positive integer into other positive
integers grows exponentially.  Solving models with adjoint matter in
DLCQ therefore requires working with exponential algorithms.

In practice, the set of states and the elements of the Hamiltonian
matrix can be generated with the aid of a computer program.  The
number of states in each sector labeled by $K$ is summarized in table
\ref{table1}.
\begin{table}
\centerline{\begin{tabular}{|c|c|c|c|c|c|c|c|c|c|c|c|c|c|c|c|} \hline
K & 11 & 12 & 13 & 14 & 15 & 16 & 17 & 18 & 19 & 20 & 21 & 22 & 23 & 24 & 25 \\ \hline
Dim & 18 &28  &40 &58  & 93 &141  & 210 & 318 & 492 &762 & 1169 & 1791 & 2786 &4338   & 6712 \\ \hline
\end{tabular}}
\caption{Number of states as a function of K in adjoint fermion model using anti-periodic boundary conditions \label{table1}}
\end{table}

The Hamiltonian preserves the number of oscillators modulo 2. The
states with odd oscillator number correspond to fermions and arise for
odd $K$. For even $K$ we necessarily have an even number of
oscillators; therefore, these states describe the bosonic part of the
spectrum. The Hamiltonian also preserves a $Z_2$ symmetry
corresponding to reversing the order in which the modes act on the
vacuum:
$$\tr [B(-n_1) B(-n_2) \ldots B(-n_l)] | 0 \rangle\leftrightarrow
\tr [B(-n_l) B(-n_{l-1}) \ldots B(-1)] | 0 \rangle
\ ,$$
so that the Hilbert space decomposes into sectors odd and even under
the action of this $Z_2$ group.

\subsection{Numerical results and extrapolations}

\begin{figure}
\centerline{\psfig{file=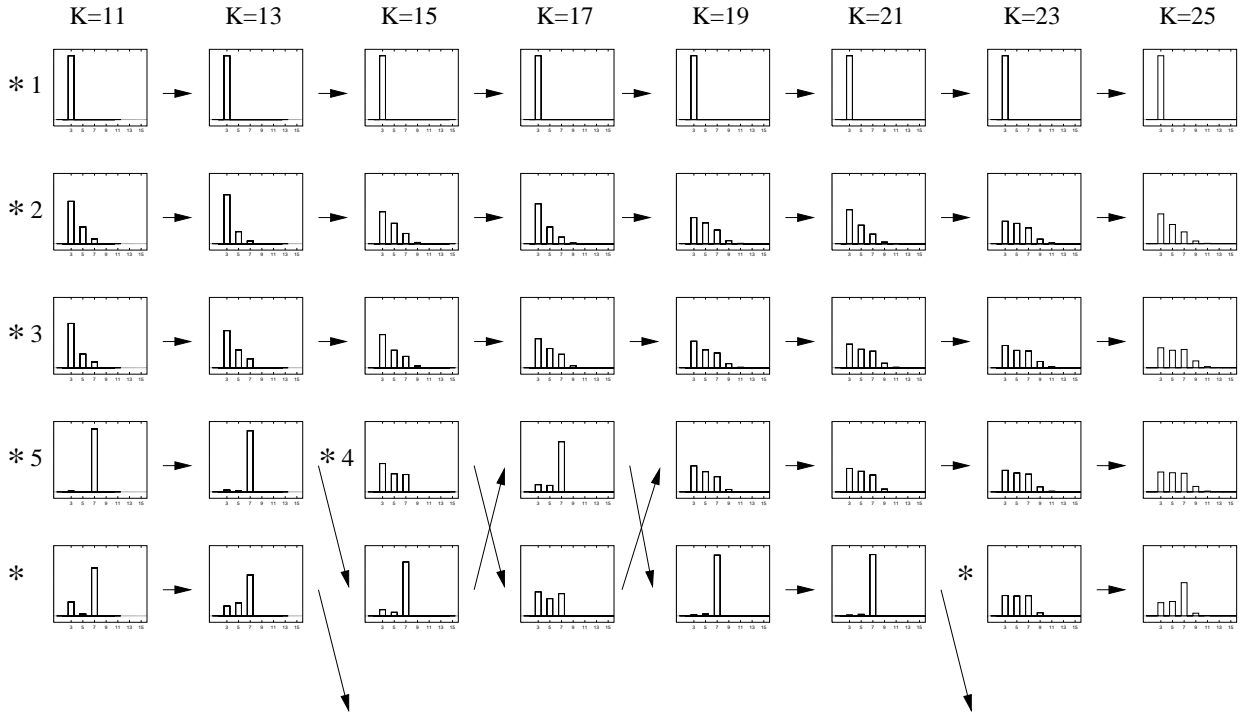,width=\hsize}}
\caption{Probability distributions in bit number space of the
low-lying mass eigenstates of the light-cone Hamiltonian in the
fermionic $Z_2$ sector. The arrows indicate the likely tracking
pattern for these states as we vary $K$.  The symbol ``*'' indicates a
``trail head'' where a new state appears in the
spectrum.\label{wavefunction}}
\end{figure}

We will now present the results of our numerical analysis.  We
evaluated the $M^2$ matrix explicitly and computed the first several
eigenvalues in each of the $Z_2$ sectors for both bosons and fermions.
We are interested in tracking the mass squared of a given state as we
vary $K$.  For this purpose we found it useful to plot the probability
that a given state has $n$ bits, which is encoded in its wavefunction,
and to track these probabilities as we increase $K$.  As an example,
we illustrate in figure \ref{wavefunction} the probabilities of
various bit numbers for each of the low-lying eigenstates. One of the
features visible in figure \ref{wavefunction} is the existence of
states which are sharply peaked in bit number distributions (e.g.\
states 1 and 5 in figure \ref{wavefunction}).  These states can be
readily distinguished from the rest of the states which are
superpositions of various bit number sectors. In figure
\ref{wavefunction} we indicate by arrows the patterns with which
states are tracked as we vary $K$. These choices are based on
continuity in the shape of the distribution and of the
eigenvalues. Although there is some element of guesswork in making
such assignments, the steady pattern we observe in the shape of the
distribution provides us with confidence that we are tracking the
states correctly.  As $K$ increases, we also observe evidence for new
tracks of states appearing in the spectrum.  This is to be expected
since the dimension of the Hilbert space increases rapidly with $K$.
We indicate the likely ``trail heads'' of these tracks with ``*'' in
figure \ref{wavefunction}.

\begin{figure}
\centerline{
\psfig{file=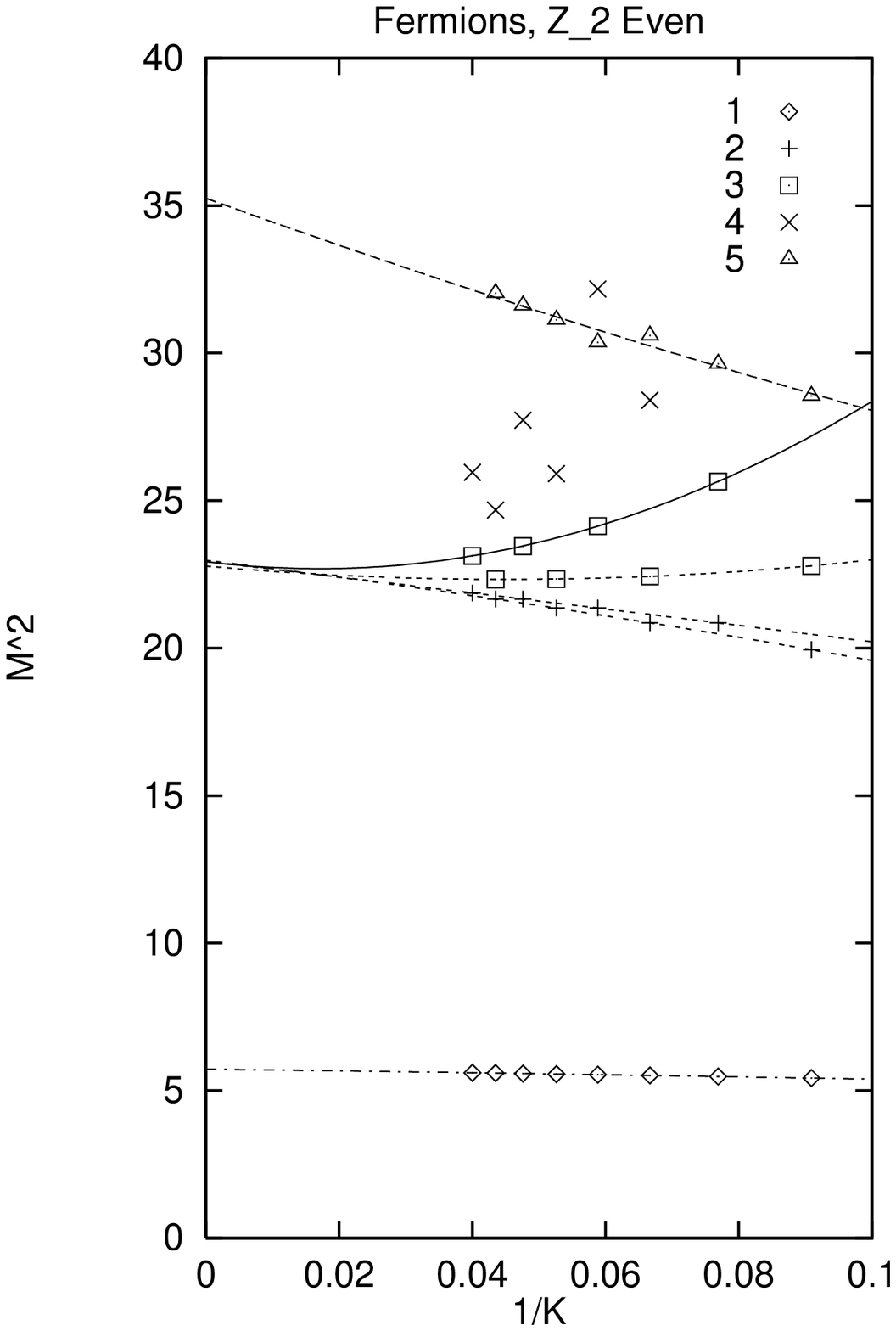,width=3in}
\psfig{file=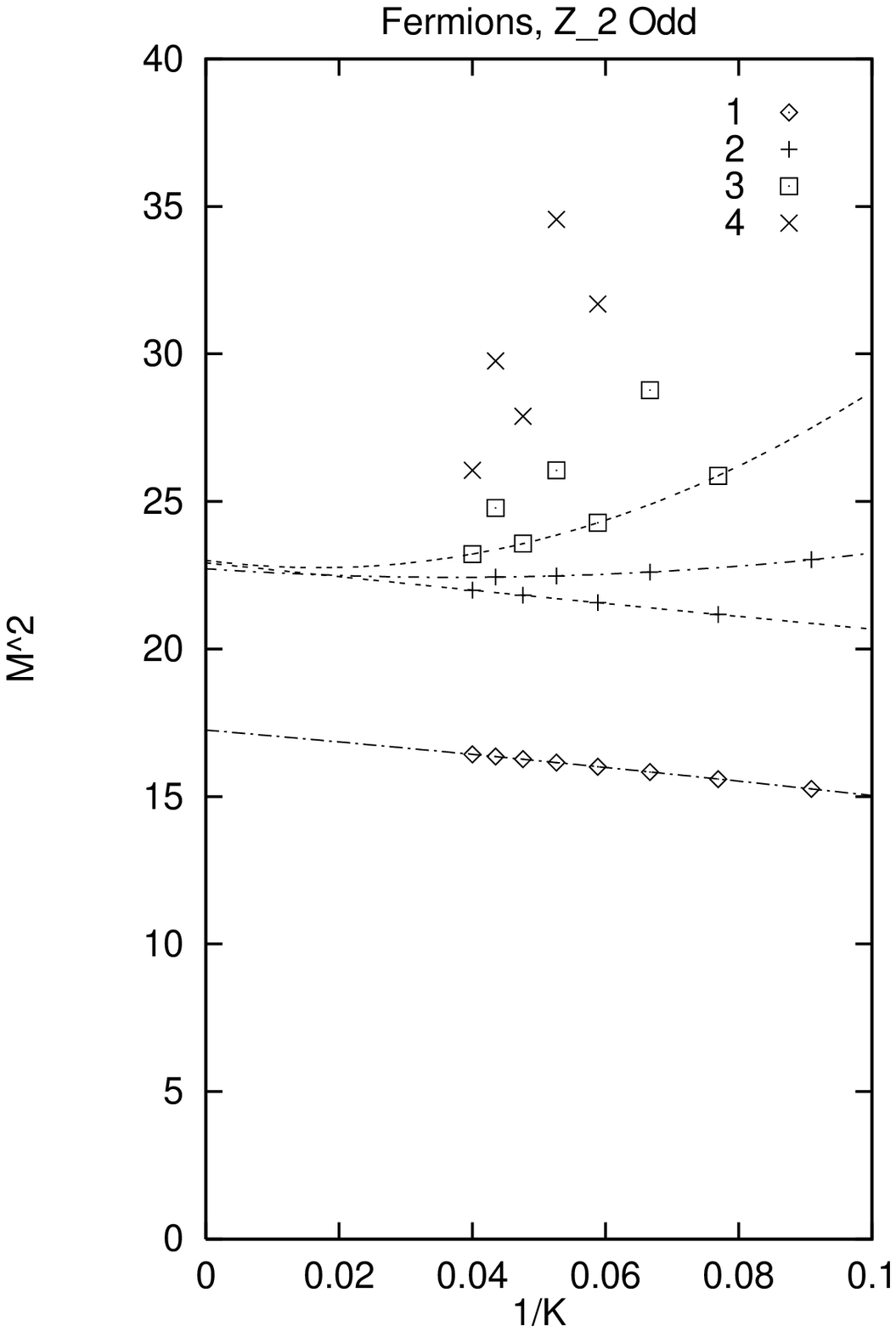,width=3in}
}
\centerline{
\psfig{file=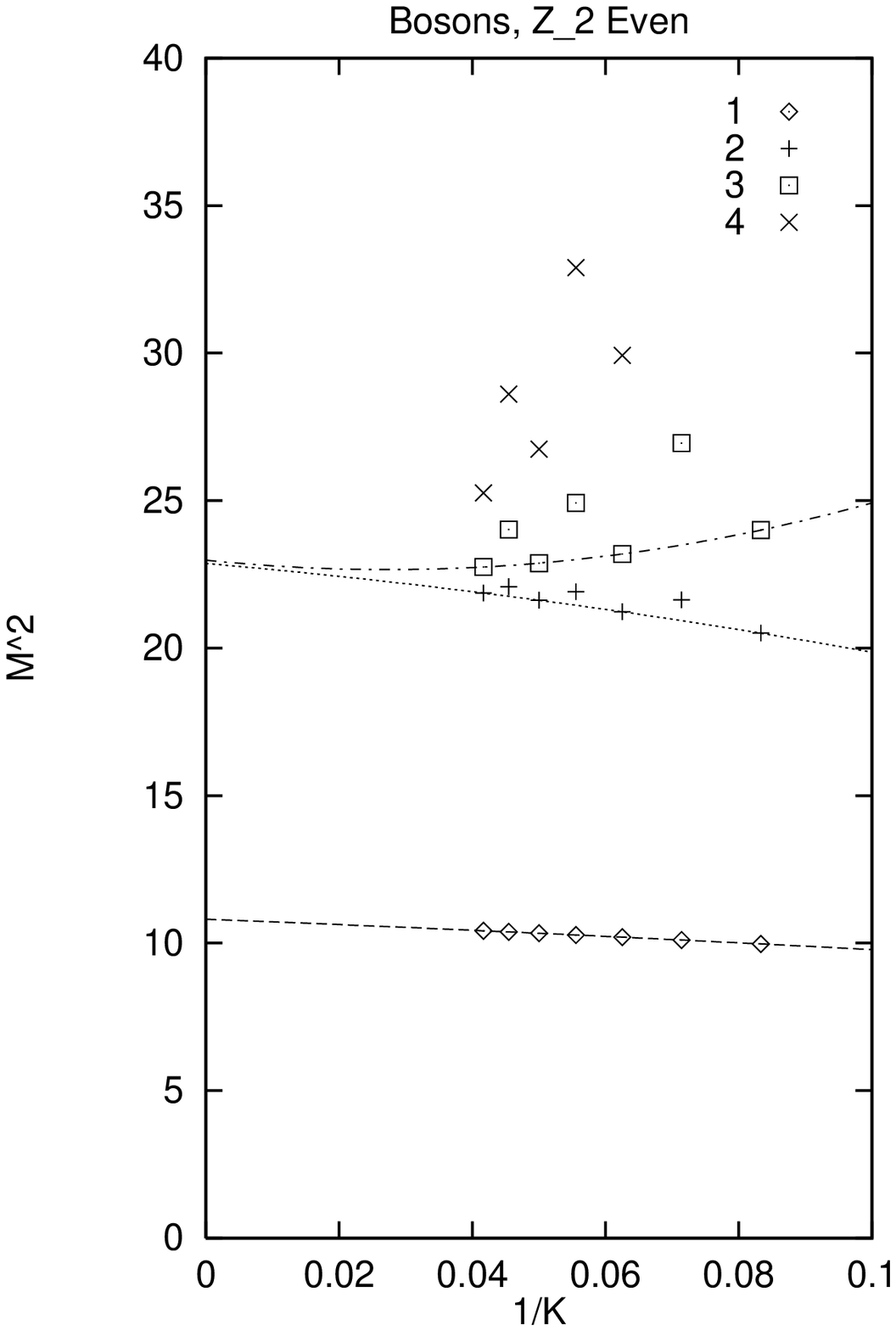,width=3in}
\psfig{file=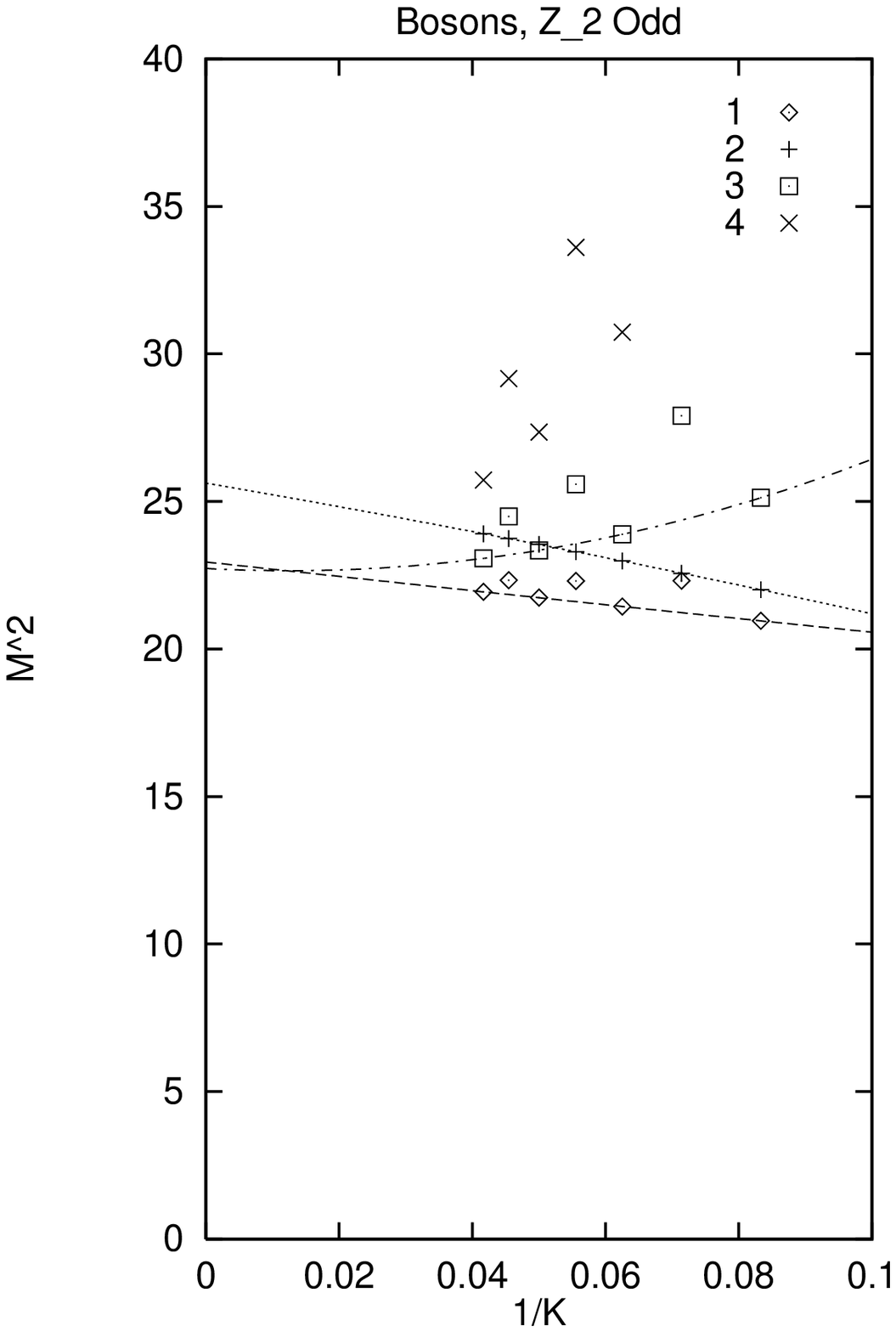,width=3in}
}
\caption{$M^2$ eigenvalues for $m=0$ as a function of $1/K$. In each
of the 4 sectors ($Z_2$ even and $Z_2$ odd for bosons and fermions) we
exhibit extrapolations of the lightest states that are pure in bit
number, and also graph some of the lightest states that appear to
converge to the continuum.
\label{fig1}}
\end{figure}

We can now follow the arrows in figure \ref{wavefunction} and plot
$M^2$ as a function of $K$. We summarize this data in figure
\ref{fig1} where we plot $M^2$ against $1/K$ for each of the $Z_2$
sectors for both bosons and fermions.

Several comments are in order regarding the data contained in figure
\ref{fig1}.  First of all, in all but the bosonic $Z_2$ odd sector,
the lowest eigenvalue appears to be well separated from the rest of
the spectrum, and depends relatively smoothly on $K$.  We studied
convergence to the large $K$ limit by performing a best fit to a curve
of the form
\begin{equation}
a_0 + a_1 \left({1 \over K}\right)
+ a_2  \left({1 \over  K } \right)^2.
\label{fit}
\end{equation}
We included the quadratic term in the fit to account for the curvature
in the data.  Thus, $a_0$ gives the extrapolated value of the mass
squared.  Our extrapolation indicates that the lightest state is a
fermion with $M_{F1}^2 = 5.7$, followed by a boson with $M_{B1}^2 =
10.8$, followed by a fermion with $M_{F2}^2 = 17.3$ (in units of $g^2
N/\pi$).  These states are approximate eigenstates of the number of
bits with eigenvalues $3$, $2$ and $5$ respectively.  These results
are in good agreement with the extrapolations performed in
\cite{BDK,KresimirIgor}.  The best fit curves also indicate that the
lightest states have converged fairly well, although for a linear
extrapolation to be justifiable, $K$ must be of order $10^2$.

The states which are not approximately pure in bit number (e.g.\
states 2, 3, and 4 in figure \ref{wavefunction}) behave somewhat
differently.  These states oscillate strongly in $K$ with a period of
4. We will have more to say about these oscillations in the following
subsection but, for the time being, simply note that they get smaller
with increasing $K$. Figure \ref{fig1} gives an indication that these
levels will ultimately converge at large $K$.  It is not sensible to
fit a curve of the form (\ref{fit}) to such a wildly oscillating
data. As an alternative, we adopt the procedure where we fit
(\ref{fit}) to the valleys and the peaks of the data separately.
Unfortunately, this cuts in half the amount of data used in each
extrapolation.  Since (\ref{fit}) is a 3 parameter fit, we performed
the extrapolation only in cases where at least 4 data points are
available.

As we illustrate in figure \ref{fig1}, these states show an indication
that they are converging toward roughly the same mass in the large $K$
limit. What makes this particularly interesting is the fact that the
extrapolated value based on the fit (\ref{fit}) is approximately
$M_G^2 = 22.9$.  The states becoming degenerate at a particular value
of $M_G^2$ is suggestive of the onset of continuous spectrum.  Notice
that $M_G^2=22.9$ equals $4 M_{F1}^2$, which is where one expects the
first band of continuum corresponding to the free two body spectrum of
particles of mass $M_{F1}$.

One could in principle perform a 3 parameter fit to 3 data points.
These fits also give extrapolations close to 22.9. The extent to which
an extrapolation misses this mark increases as we go up in energy
levels. However, since higher levels start at trail heads with higher
values of $K$, it is necessary to go to higher values of $K$ to
achieve the same degree of convergence.  We expect these higher levels
to converge to a mass-squared of 22.9 when the calculation is pushed
to sufficiently high $K$ to allow for a reliable extrapolation.  The
data illustrated in figure \ref{fig1} is quite suggestive of such
degeneracies at large $K$.

A picture that appears to be emerging from these observations is the
following. $M_{F1}$, $M_{B1}$, and $M_{F2}$ are the masses of the
lightest particles to which the single trace states (\ref{singlets})
dissociate in the deconfinement limit.  The states piling up at $M_G$
suggest a continuous two-body spectrum of $|F1\rangle$.

There is evidence for other ``single-particle'' states buried in the
continuum.  A clear example is the bosonic $Z_2$ odd state (state 2 in
figure \ref{fig1}) which, to a good approximation, consists of 4 bits
and has the extrapolated mass-squared equal to $25.6$.  Furthermore,
we find evidence for a very pure 7-bit state (state 5 in figure
\ref{fig1}) of extrapolated mass-squared $35.3$ in the fermionic $Z_2$
even sector. The masses of these ``pure'' states appear to vary
smoothly with $K$ similarly to those of the ``single-particle'' states
$|F1\rangle$, $|B1 \rangle$, and $|F2 \rangle$.  Thus, we speculate
that there is an infinite sequence of ``single-particle'' states.  At
least the first few such states are distinguished by their purity in
the number of bits (the ``multi-particle'' states tend to be far from
being eigenstates of the number of bits).
 
It follows that there should also be other two-body thresholds at
mass-squared $(M_{F1} + M_{B1})^2 = 32.2$, $(M_{F1}+M_{F2})^2 = 42.8$,
$(M_{B1}+M_{F2})^2 = 55.4$, etc.  Perhaps these could be detected by a
sufficiently detailed examination of the spectrum.

One final point we wish to emphasize is the fact that the continuum at
$M_G^2=22.9$ appears to exist both in the bosonic and the fermionic
sectors.  The interpretation of these states as the two-body continuum
coming from the $|F1\rangle$ particle suggests that these states
should be bosonic, $|F1 \rangle \otimes |F1 \rangle$. How then should
we interpret the continuum at $M_G=22.9$ in the fermionic sector?
Recalling a similar issue of statistics in the case of adjoint v.s.\
$N$-flavor fundamental correspondence suggests the following
explanation.  The states in the fermionic continuum must correspond to
a state of the form
$$|F1\rangle \otimes |F1\rangle \otimes |\Xi\rangle$$
where $|\Xi\rangle$ is a companion fermionic state arising from the
topological sector of the theory, which is otherwise decoupled from
the dynamics.

\subsection{Oscillations}

So far, our evidence for the appearance of a continuum of states at
$M^2 = 4 M_{F1}^2$ has been based on numerical extrapolations. In this
subsection, we will present stronger evidence by studying the pattern
of oscillations exhibited by these states as they converge toward the
large $K$ limit.

Oscillations similar to the ones we illustrated in figure \ref{fig1}
arise in the DLCQ spectrum of a pair of free particles of mass $m$.
For a finite $K$, the spectrum is given by
$$ M^2 = m^2 K\left({1 \over n} + {1 \over {K-n}}\right)$$
where $n$ and $K-n$ are the the numbers of units of momentum carried
by the individual particles, and $1 \le n \le K-1$. This spectrum
oscillates in $K$. We will show that the oscillations seen in figure
\ref{fig1} are due to a similar mechanism.

\begin{figure}
\centerline{\parbox[t]{3in}{
\psfig{file=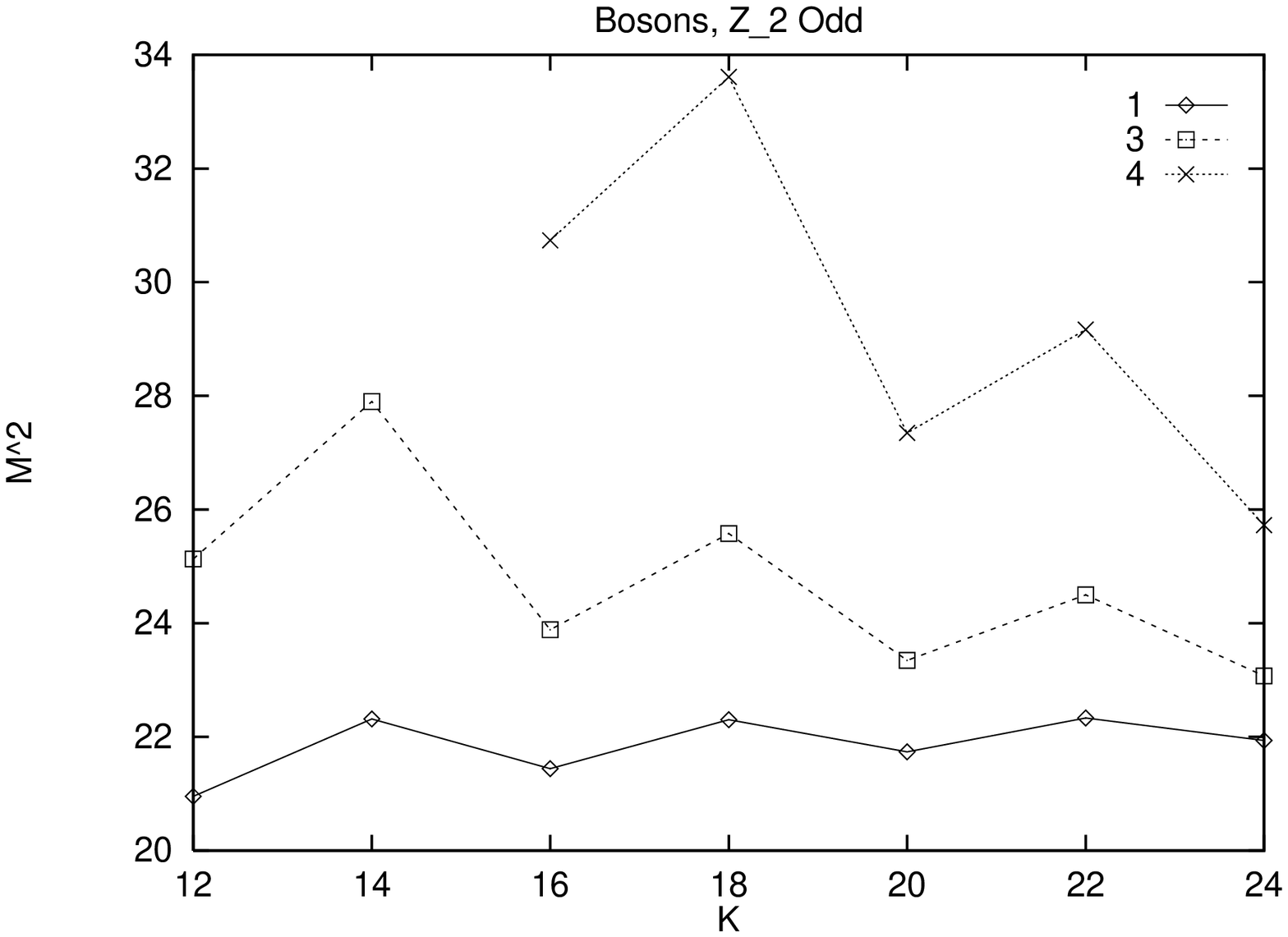,width=\hsize}
\caption{The first three states in the bosonic $Z_2$ odd sector of the
adjoint fermion model that appear to be converging to the
continuum. \label{zigzag1}}} \hspace{2ex}
\parbox[t]{3in}{
\psfig{file=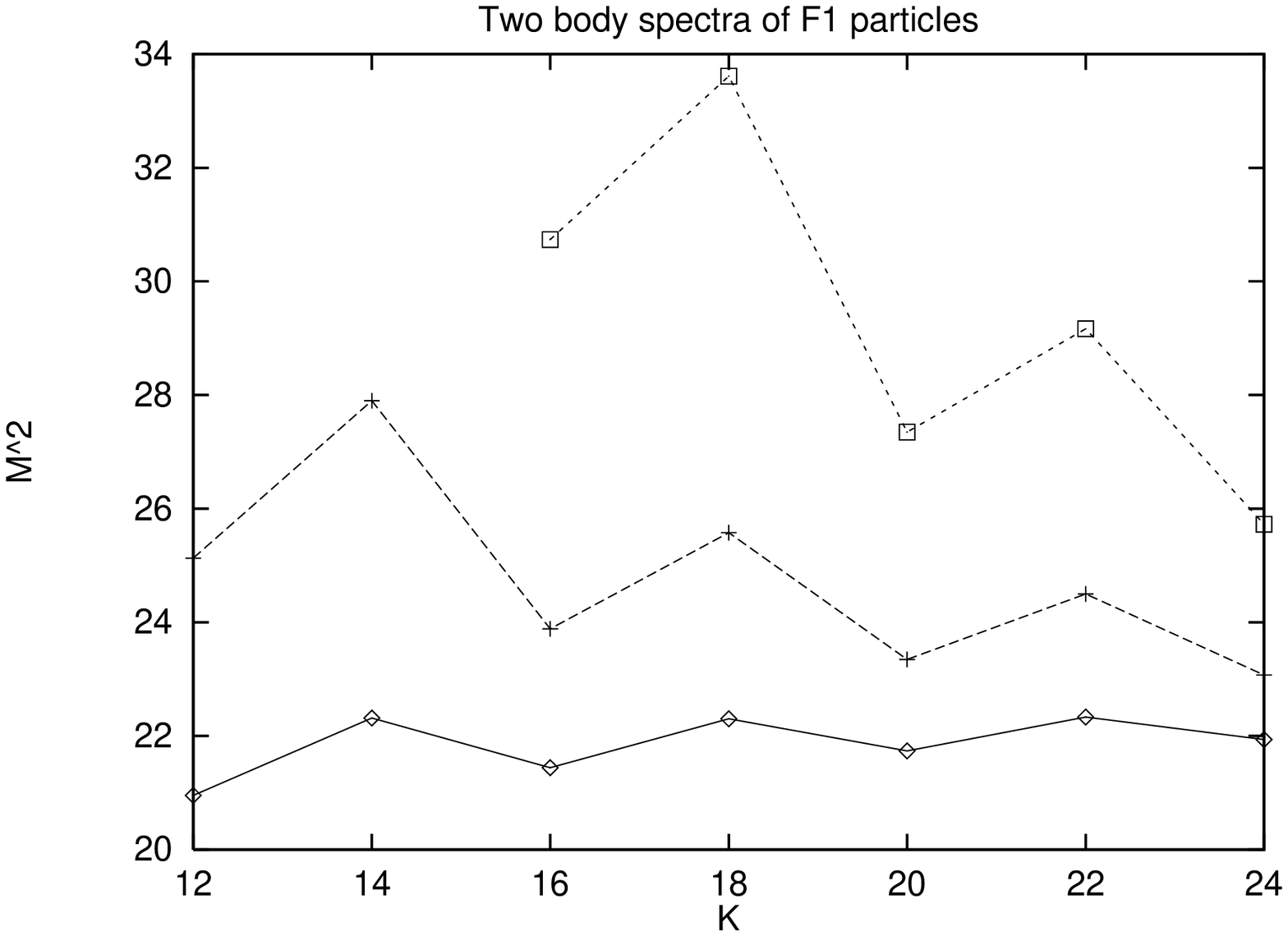,width=\hsize}
\caption{ The spectrum of a pair of non-interacting F1 particles with
total momentum $K$. \label{zigzag2}}}}
\end{figure}

Let us focus our attention on the oscillating states in the bosonic
$Z_2$ odd sector from figure \ref{fig1}.  For the sake of
illustration, we plot these states again in figure \ref{zigzag1}.  Our
claim is that these states arise as the free two-body spectrum of
$|F1\rangle$ particles. Since each of these particles is a composite
state, whose mass is determined with a finite resolution, the correct
formula for a finite $K$ is
\begin{equation}
M^2 = K \left( {M^2_{F1}(n) \over n} + {M^2_{F1}(K-n) \over (K-n)} \right)
\ ,\label{2body}
\end{equation}
where $n$ is an odd integer $1 \le n \le K-1$, and $M_{F1}(n)$ is the
mass of the $|F1\rangle$ in the $K=n$ sector. We illustrate the
spectrum determined using (\ref{2body}) in figure
\ref{zigzag2}. (Since $|F1\rangle$ is a fermion, we only keep the
states which are antisymmetric with respect to their exchange.)

Figures \ref{zigzag1} and \ref{zigzag2} are generated using completely
independent methods. It is therefore quite remarkable that the
resulting plots are {\em identical}. This can be easily verified by
laying one on top of the other. Thus, the identification of bosonic
$Z_2$ odd states as non-interacting two-body states of $|F1\rangle$
appears to be exact even for finite $K$.

\begin{figure}
\centerline{\parbox[t]{3in}{
\psfig{file=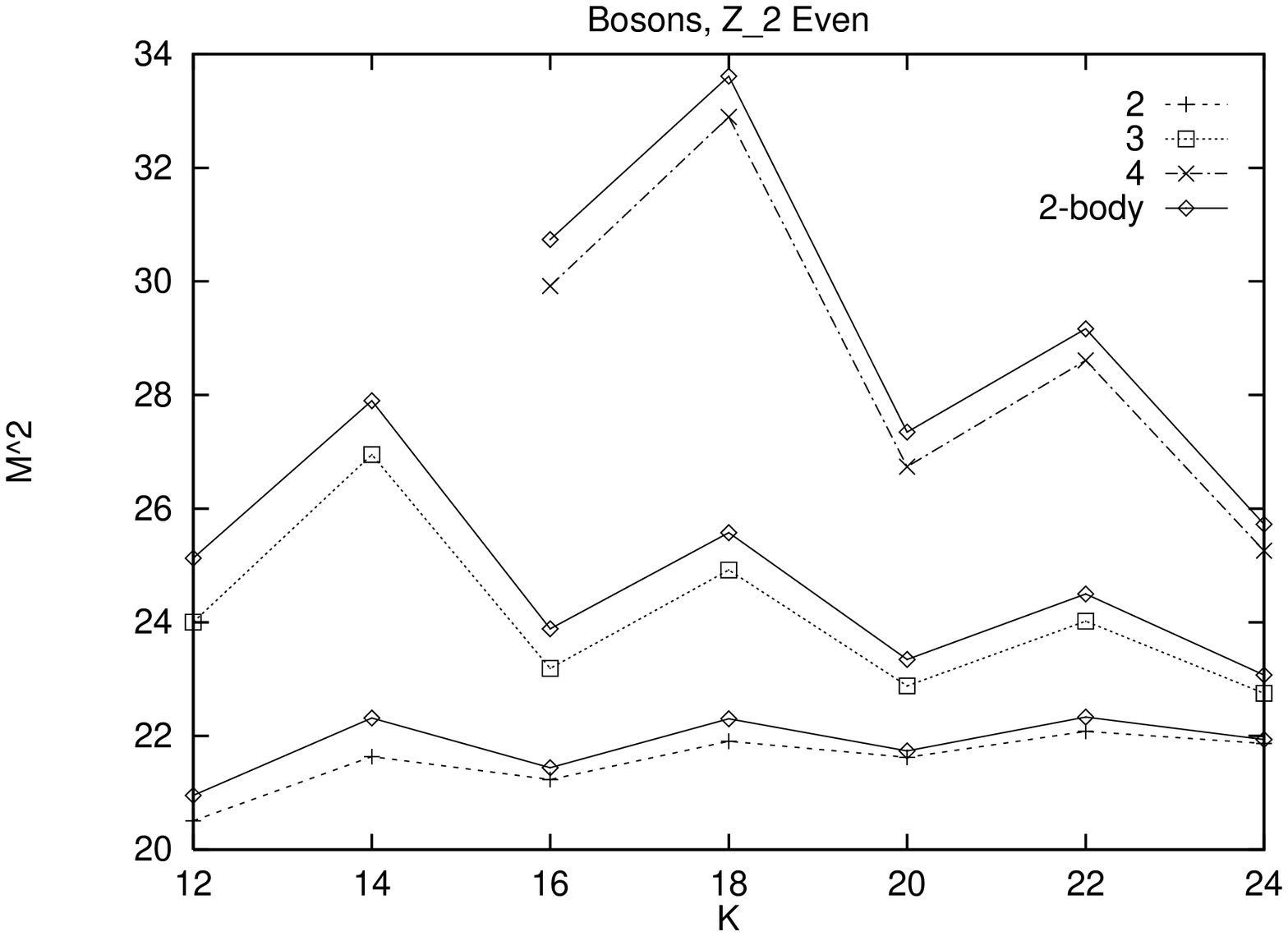,width=\hsize}
\caption{Dashed lines label the first three states in the bosonic
$Z_2$ even sector of the adjoint fermion model that appear to be
converging to the continuum. The solid line is the spectrum of a
non-interacting pair of F1 particles. \label{zigzag3}}}
\hspace{2ex}
\parbox[t]{3in}{
\psfig{file=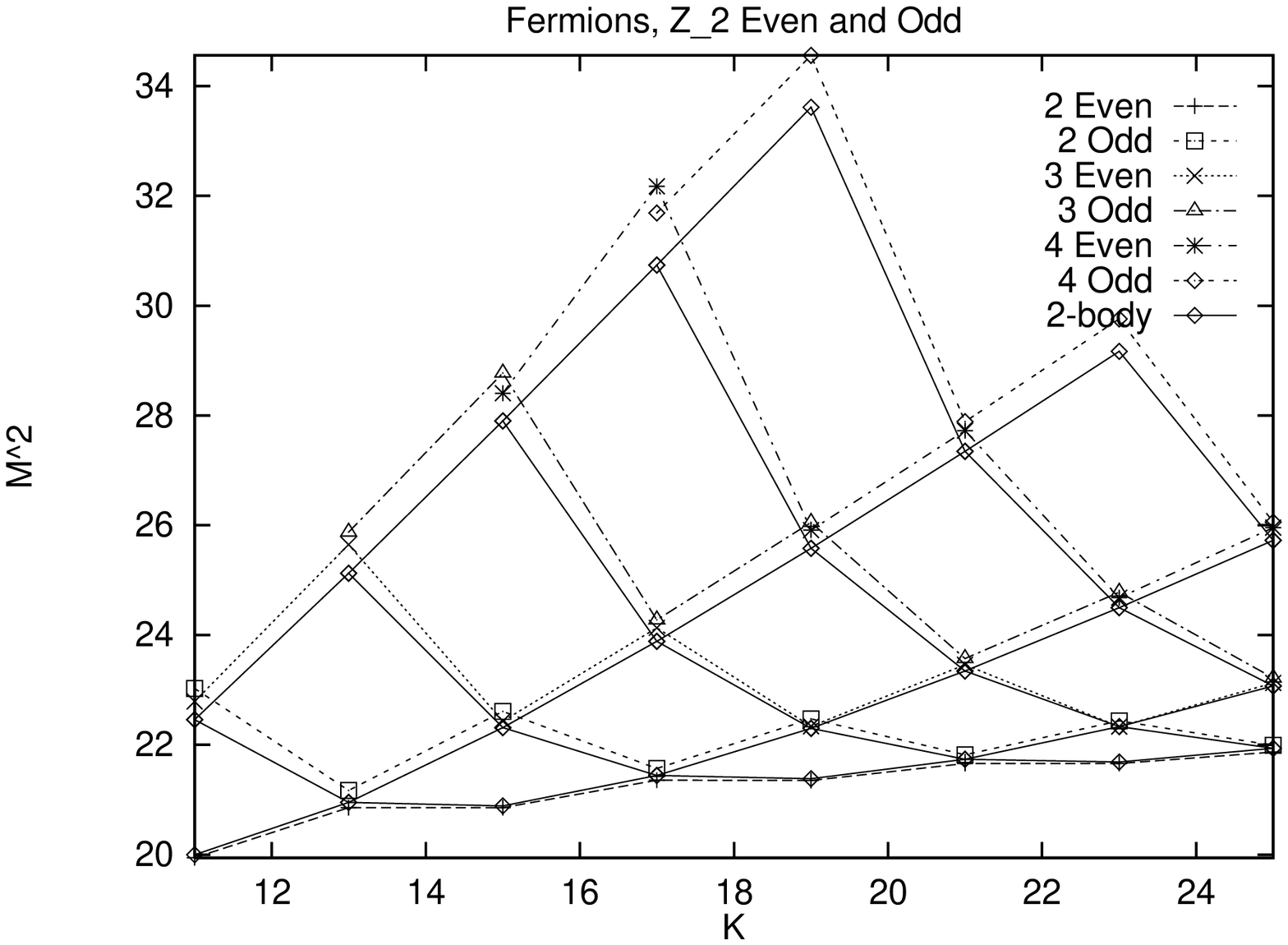,width=\hsize}
\caption{Dashed lines label the first three states for both the
fermionic $Z_2$ even and the fermionic $Z_2$ odd sectors of the
adjoint fermion model that appear to be converging to the
continuum. The solid line is the spectrum of a non-interacting pair of
F1 particles.  \label{zigzag4}}}}
\end{figure}

Having found such a remarkable structure in the $Z_2$ odd sector of
the bosonic spectrum, it is natural to expect a similar situation to
hold in the $Z_2$ even sector. However, the correspondence here is not
exact.  In figure \ref{zigzag3}, we illustrate both the bosonic $Z_2$
spectrum from previous section and the expectation based on
(\ref{2body}). What we seem to be finding here is that (\ref{2body})
captures the qualitative features of the oscillations of the $Z_2$
even sector states, but the DLCQ data contains additional ``noise.''
We will speculate on the source of this noise at the end of this
section.  Empirically, we find that the noise decays as $1/K$ in the
large $K$ limit. Therefore, despite the fact that the correspondence
is not exact at finite $K$, we believe that the pattern of oscillation
seen in the bosonic $Z_2$ even sector is also consistent with the
picture that a continuum is formed in the large $K$ limit.

There are additional subtleties in attempting to extend this picture
to the states in the fermionic sector.  As was described at the end of
the previous subsection, the only way one can make a fermionic state
out of a pair of $|F1\rangle$ particles is to introduce states from
the topological sector of the theory carrying fermionic statistics. In
DLCQ with anti-periodic boundary conditions, such a state cannot carry
zero light-cone momentum, as these states are required to carry
momentum in half-integer units. The closest we can get to the
situation we hope to describe is to distribute $K$ units of momentum
according to $(|F\rangle, |F\rangle, |\Xi\rangle) = (n, K-n-1, 1)$. It
should be stressed, however, that $|\Xi\rangle$ is not really a free
particle, and it is not a priori clear how to generalize (\ref{2body})
taking the ``topological sector'' into account. A rough guess is to
take
\begin{equation}
M^2 = (K-1) \left( {M^2_{F1}(n) \over n} + 
{M^2_{F1}(K-n-1) \over (K-n-1)} \right).
\label{2body.f}
\end{equation}
The data from the fermionic sector of the adjoint fermion model
suggests the presence of two-particle states which are both symmetric
and antisymmetric under their exchange.  This is puzzling in light of
the fact that $|F1\rangle$'s constitute a pair of identical
fermions. Perhaps $|\Xi \rangle$ is binding with one of the
$|F1\rangle$'s so that the resulting pair of constituents are no
longer identical. Here we have included the symmetric wavefunctions
for the sake of comparison with the DLCQ data.

Although (\ref{2body.f}) is admittedly ad-hoc, it appears to capture
the general structure of the oscillations seen in the spectrum
computed for the adjoint fermion model, as we illustrate in figure
\ref{zigzag4}.  Again, we find qualitative agreement accompanied by
some ``noise'' which decays as $1/K$. This time, however, there is a
natural suspect for the culprit responsible for the noise.  The
topological sector plays an important role in assigning appropriate
statistics for the states in this sector. By using anti-periodic
boundary conditions, however, we were forced to mutilate the structure
of the topological sector by forcing it to carry small but finite
momenta. This effect is not properly accounted for in
(\ref{2body.f}). The amount of error in momentum we introduced is of
order $1/K$, and this is indeed the magnitude of the noise we see in
the spectrum.

This also suggests the probable cause of the noise in the bosonic
$Z_2$ even sector.  The structure of these states might very well be
of the form
$$|F1 \rangle \otimes |F1 \rangle \otimes |\Omega \rangle$$
where $|\Omega \rangle$ is a topological state carrying even fermion
numbers. Even in theories with only fermionic matter, such a state can
arise easily from bilinears. The noise in this sector may be due to
the fact that (\ref{2body}) does not account for the presence of the
topological sector of this type.  In the large $K$ limit, however, we
would expect all dynamics in the topological sector to decouple. It is
therefore satisfying to find that the noise decays as $1/K$ in the
large $K$ limit.

Although in general the correspondence between the spectra derived
from (\ref{2body}) and (\ref{2body.f}) with the DLCQ spectrum is not
as precise as what we found in the bosonic $Z_2$ odd sector, the
qualitative agreement, and the fact that the discrepancy shrinks with
increasing $K$, provides a strong indication that the oscillations
seen in the DLCQ spectrum are a signature of states forming a
continuum. Furthermore, we feel that the small discrepancy is actually
probing the topological sector of the theory. It would be extremely
interesting to understand this structure from first principles.

\section{Conclusions}

By performing explicit DLCQ analysis of QCD coupled to adjoint
fermions, we found isolated ``single-particle'' states $|F1 \rangle$,
$|B1 \rangle$ and $|F2 \rangle $ at mass-squared equal to 5.7, 10.8,
and 17.3 respectively (in units of $g^2 N/\pi$).  These states are
approximate eigenstates of the number of bits with eigenvalues $3$,
$2$ and $5$ respectively. In addition, we found an indication that a
continuum of states is appearing at $M_G^2 = 22.9$ in both the bosonic
and the fermionic sector. The fact that $M_G^2 = 4M_{F1}^2$ suggests
that these states are a two-particle continuum built out of the
$|F1\rangle$'s. To account for the statistics, the states are
interpreted to be of the form $|F1\rangle \otimes |F1 \rangle$ or $|F1
\rangle \otimes |F1 \rangle \otimes |\Xi \rangle$, where $|\Xi
\rangle$ describes a fermionic state in the topological sector of the
theory, which is otherwise decoupled from the dynamics.  The existence
of fermionic states converging to the continuum at $M_G^2 = 22.9$ thus
provides some numerical evidence for the ``direct sum of tensor
products'' structure of the Hilbert space,
$${\cal H} = \sum_{s,s'} \oplus ( 
{\cal H}_s^c \otimes
{\cal H}_{s,s'}^{GI} \otimes
{\cal H}_{s'}^c 
)
$$
suggested in \cite{KutasovSchwimmer}.  The oscillatory behavior of
$M^2$ for the states converging to the continuum can be understood
exactly (at least for the bosonic $Z_2$ odd states) in terms of the
spectrum of a non-interacting 2-body system,
$$
M^2 = K \left( {M^2_{F1}(n) \over n} + {M^2_{F1}(K-n) \over (K-n)} \right)
\ .$$
This provides strong support for our claim about the continuity of the
spectrum for $M^2 > 4 M_{F1}^2$.

In addition, we find evidence for other ``single-particle'' states
whose mass is higher than $M_G$. At least the first few such states
are distinguished from the continuum states by their purity in bit
number. For example, in the bosonic $Z_2$ odd sector there is a state
of mass-squared $25.6$ which is, to a high accuracy, a 4-bit state. We
are thus led to speculate that there is an infinite sequence of
``single-particle'' states.  Perhaps these states can be grouped into
one Regge trajectory of the fermions and one Regge trajectory of the
bosons.

Our analysis indicates clearly that the adjoint fermion model contains
string-like states made out of adjoint bits which dissociate in the $m
\rightarrow 0$ limit into the stable constituent ``particles.''  For
small $m$ these states can be thought of as loosely bound state of
such ``particles'' In the $m\rightarrow 0$ limit, these ``particles''
are free, as can be inferred from the threshold of the continuum.

While we have seen that the DLCQ gives convincing evidence for the
existence of constituent ``particles'' and their 2-body continua, some
puzzles about the structure of the spectrum remain.  A paradox having
to do with the state counting of this model was noted in
\cite{BK97}. Since the tension of the QCD string vanishes in the
$m\rightarrow 0$ limit, one expects to find a spectrum with Hagedorn
temperature $T_H \rightarrow 0$.  On the other hand, one expects the
spectrum of a screening theory to have $T_H = \infty$.  Since for
$m=0$ the spectrum decomposes into the more basic building blocks
(single particles), we should only count these particles as
fundamental. It is natural to expect that these particles form a
single Regge trajectory, hence they do not have an exponentially
growing density of states.  The problem is that the particles from a
single Regge trajectory, and the multiparticle states built out them,
do not have enough degeneracy to form an exponentially rising density
of states when $m$ is turned on \cite{BK97}.  Thus, it seems necessary
to take into account additional large degeneracy due to the presence
of certain topological states \cite{KutasovSchwimmer,BK97}.  It would
be interesting to see how the resolution to this apparent paradox
manifests itself in the DLCQ numerical analysis.

\section*{Acknowledgments}

We are grateful to D. Kutasov for discussions. We also acknowledge
support from Supercomputer Facility at UCSB (NSF Grant CDA96-01954).
The work of D.J.G.\ and A.H.\ was supported in part by the NSF grant
PHY94-07194.  The work of I.R.K.\ was supported in part by the DOE
grant DE-FG02-91ER40671, the NSF Presidential Young Investigator Award
PHY-9157482, and the James S.\ McDonnell Foundation grant No.\ 91-48.

\bibliographystyle{plain} \bibliography{adjoint}
\end{document}